\documentclass{article}

\usepackage{microtype}
\usepackage{graphicx}
\usepackage{subcaption}
\usepackage{booktabs}
\usepackage{hyperref}
\usepackage{enumitem}
\usepackage{tikz}
\usetikzlibrary{arrows.meta}

\usepackage[preprint]{icml2026}

\usepackage{amsmath}
\usepackage{amssymb}
\usepackage{mathtools}
\usepackage{amsthm}
\usepackage[capitalize,noabbrev]{cleveref}
\usepackage{footmisc}

\theoremstyle{plain}

\theoremstyle{definition}

\theoremstyle{remark}

\usepackage[textsize=tiny]{todonotes}

\icmltitlerunning{Bit-Exact AI Inference Verification Without Performance Tradeoffs}

\begin{document}

\twocolumn[
  \icmltitle{Bit-Exact AI Inference Verification Without Performance Tradeoffs}

  \icmlsetsymbol{equal}{*}

  \begin{icmlauthorlist}
    \icmlauthor{Naci Cankaya}{MIRI}
  \end{icmlauthorlist}

  \icmlaffiliation{MIRI}{Machine Intelligence Research Institute}

  \icmlcorrespondingauthor{Naci Cankaya}{naci@intelligence.org}

  \icmlkeywords{AI governance, verification, floating-point arithmetic, tensor cores, inference}

  \vskip 0.3in
]

\printAffiliationsAndNotice{}

\begin{abstract}
Verifying claims about AI workloads is a prerequisite for credible AI governance of covert adversaries (who comply with monitoring only when detection likelihood is high), yet the \textit{apparent} non-determinism of GPU floating-point arithmetic forces auditors to accept approximate output matches. Covert adversaries can exploit unverifiable degrees of freedom in monitored computation. Attack vectors include steganography, unreported modification of inference software, and covert computation via unreported batch elements. Empirically, we analyze how modern inference engines (vLLM, HF transformers) produce deterministic but non-invariant outputs, without needing to set performance-compromising determinism flags, if the right information is available for re-computation and no atomic functions are called in the backend. We demonstrate that such bitwise-precise re-computation does not require access to identical hardware, via a software-only emulation of LLM inference across multiple NVIDIA GPU variants. Thus, accumulated rounding errors can be an auditable \textit{signature} of the software and hardware setup used for inference, instead of a constraint on verifiability. Source code for the emulator is available \href{https://github.com/NaciCankaya/hardware_rounding_error_predictor}{\texttt{here}}, and the repository for the empirical studies can be found \href{https://github.com/NaciCankaya/Floating_point_noise_GPU_verification}{\texttt{here}}.
\end{abstract}

\section{Introduction and Related Work}
\label{sec:intro}

The motivation for this work is verification of the claimed ML computation of a covert adversary~\cite{aumann2010covert}: an adversary who will exploit gaps in a verification setup while complying only if the monitoring ensures high likelihood of detecting false claims. Such monitoring and verification of covert adversaries describes the threat model of low-trust AI governance, e.g.\ verification of an international, mutual agreement between rival nations, for restrained AI development~\cite{baker2025verifying, harack2025verification, scher2024verify, wasil2024verification}. An untrusted owner of AI compute resources (the ``prover'') can make claims about the computation performed on their devices to an outside party (the ``verifier'') who chooses claimed computations for retroactive verification. The verifier can check a claim $f(x) = y$ for self-consistency by re-running the computation on their own hardware or via a zero-knowledge proof~\cite{sun2024zkllm} in which the prover demonstrates correctness without revealing the input.

AI accelerators rely on parallel, low-precision computation. This work focuses on the resulting challenge to retroactive verification: rounding errors in $f(x)$ from non-associative summation across parallel reductions~\cite{goldberg1991float, shanmugavelu2024fpna, yuan2025fp32}. In contemporary inference of AI algorithms, this is perceived as ``non-deterministic'' noise in outputs, and can create plausible deniability in a verification setup:~\cite{rinberg2025exfiltration} describe a steganographic attack vector exploiting unverifiable degrees of freedom on ML outputs for covert communication. There is increasing interest in network taps for ensuring only monitored communication in and out of untrusted servers~\cite{cankaya2025fingerprinting}, but steganography hides even in fully monitored traffic. Furthermore, unverifiable rounding errors may enable covert modification of inference and training software, in an attempt to hide additional computation on a monitored server (e.g.\ via unreported batch elements and kernel optimization). Statistical verification schemes~\cite{karvonen2025difr, rinberg2025exfiltration} can upper-bound the covert bandwidth available to an adversary but cannot close it entirely. Zero-knowledge proofs on LLMs have made significant progress in terms of performance, but require determinism as a precondition~\cite{sun2024zkllm}. Prior work has achieved output invariance across batch sizes, using custom kernels, though at the cost of $\approx20\%$ throughput loss~\cite{thinkingmachines2024}. 
Deepseek-V4's technical report goes into detail on batch-invariant and deterministic kernel libraries~\cite{deepseek2026v4}. The authors claim minimal overhead across the kernel suite, and negligible overhead specifically for batch-invariant decoding.

Without modifying inference engines with invariance-enforcing kernels\footnote{All experiments used unmodified HuggingFace Transformers and vLLM. Software versions were logged for every experiment. Version sensitivity is itself one of the configuration factors we characterize in Sections~\ref{sec:rootcauses} and~\ref{sec:empirical}. Representative stacks include vLLM~0.11.2 with PyTorch~2.9.x/CUDA~12.8, and HF Transformers~4.57.3 with flash-attn~2.8.3.}, we demonstrate bitwise verifiability of activation tensors:
\textbf{A)}~We investigate the root causes of differing rounding errors across software and hardware conditions, and distinguish true non-determinism (atomic functions) from non-invariance (deterministic, but different reduction trees) (Section~\ref{sec:rootcauses}).
\textbf{B)}~Experimentally, we demonstrate deterministic outputs, provided the right conditions are fixed (Section~\ref{sec:empirical}).
\textbf{C)}~We predict the precise outputs of LLM inference computed on different generations of NVIDIA accelerators (L40, L40S, A100, H100). For this, we use a software emulation --- running entirely on CPU --- that models every rounding decision of the GPU hardware and software stack (Section~\ref{sec:emulator}).

We conclude that modern AI inference engines produce deterministic results, which are bitwise verifiable provided the key factors are known by a verifier, which we isolate and list in Section~\ref{sec:governance}. These factors are predominantly static at inference time (hardware setup, software versions, model weights, parallelism topology), with the exception of dynamic batch size. Recording the concurrent batch size is one additional integer per forward pass to track --- negligible overhead for the prover. Reporting this information in addition to outputs removes any plausible deniability in output numerics, collapsing verification into a simple pass/fail. Notably, the numeric sensitivity of outputs is a high-fidelity \textit{signal} for those reported properties, turning non-associativity from a noise source to a fingerprint of the prover's hardware and software stack.

\section{Root Causes of Apparent Non-Determinism}
\label{sec:rootcauses}

Modern AI inference and training make heavy use of parallel computation. We distinguish two classes of arithmetic operations in transformer inference: \textbf{element-wise} and \textbf{reductions}.\footnote{Parallel prefix scans, used in state-space models such as Mamba~\cite{gu2023mamba}, are a distinct topology class governed by the same non-associativity mechanism but not strictly a reduction. We focus on transformer inference in this work.} Both can be sources of apparent non-determinism, for different reasons. 

\subsection{Element-Wise Operations}
\textbf{Element-wise} operations executed by dedicated hardware acceleration can use non-standard approximations.\footnote{The IEEE~754 standard~\cite{ieee754_2019} specifies correctly-rounded results for basic arithmetic operations but does not standardize transcendental functions, leaving implementations free to approximate~\cite{goldberg1991float}.} This is the case for some select operations on NVIDIA accelerators~\cite{nvidia_ptx_isa}, and we found that accurately predicting the outputs of transformer FFN and attention blocks requires modeling special function units (SFUs) for exponentials, reciprocals, and square roots. Software is also a source of discrepancy: trigonometric functions in RoPE, for example, depend on the specific math library used (e.g., CUDA libdevice~\cite{nvidia_libdevice} vs.\ CPU libm), and different library versions may produce different results for a fraction of inputs~\cite{nvidia_cuda_release_notes, redhat_libm_versions}. Fast-math approximations are a well-known issue for reproducibility: DeepSeek-V4's TileLang defaults to strict IEEE 754 rounding for element-wise operations, treating \texttt{fast-math} as opt-in~\cite{deepseek2026v4}.

\subsection{Reductions}
\textbf{Reductions} sum multiple parallel elements into one. Floating-point addition is non-associative: $a+(b+c) \neq (a+b)+c$. Each addition rounds, and the rounding depends on the relative magnitudes of the operands~\cite{goldberg1991float}.

\begin{figure}
\centering
\tikzset{
  op/.style={circle,draw,inner sep=0pt,minimum size=3.8mm,font=\scriptsize},
  inp/.style={draw,rectangle,inner sep=0pt,minimum width=4.2mm,minimum height=3.5mm,font=\scriptsize},
  arr/.style={-,thin}
}
\begin{subfigure}[b]{0.48\linewidth}
\centering
\begin{tikzpicture}
\node[inp] (c)  at (0,0)    {$c$};
\node[inp] (p0) at (0.55,0)  {$p_0$};
\node[op]  (a1) at (0.275,-0.65) {$+$};
\node[inp] (p1) at (0.825,-0.65) {$p_1$};
\node[op]  (a2) at (0.55,-1.3)  {$+$};
\node[inp] (p2) at (1.1,-1.3)  {$p_2$};
\node[op]  (a3) at (0.825,-1.95) {$+$};
\node[inp] (p3) at (1.375,-1.95) {$p_3$};
\node[op]  (a4) at (1.1,-2.6)  {$+$};
\draw[arr] (c)  -- (a1); \draw[arr] (p0) -- (a1);
\draw[arr] (a1) -- (a2); \draw[arr] (p1) -- (a2);
\draw[arr] (a2) -- (a3); \draw[arr] (p2) -- (a3);
\draw[arr] (a3) -- (a4); \draw[arr] (p3) -- (a4);
\end{tikzpicture}
\caption{Sequential}
\end{subfigure}%
\hfill
\begin{subfigure}[b]{0.48\linewidth}
\centering
\begin{tikzpicture}
\foreach \i in {0,1,2,3,4,5,6,7} {
  \node[inp] (p\i) at (\i*0.45,0) {$p_\i$};
}
\node[op] (l1a) at (0.225,-0.65) {$+$};
\node[op] (l1b) at (1.125,-0.65) {$+$};
\node[op] (l1c) at (2.025,-0.65) {$+$};
\node[op] (l1d) at (2.925,-0.65) {$+$};
\draw[arr] (p0) -- (l1a); \draw[arr] (p1) -- (l1a);
\draw[arr] (p2) -- (l1b); \draw[arr] (p3) -- (l1b);
\draw[arr] (p4) -- (l1c); \draw[arr] (p5) -- (l1c);
\draw[arr] (p6) -- (l1d); \draw[arr] (p7) -- (l1d);
\node[inp] (c)   at (0.225,-1.3) {$c$};
\node[op]  (l2a) at (0.675,-1.3) {$+$};
\node[op]  (l2b) at (2.475,-1.3) {$+$};
\draw[arr] (l1a) -- (l2a); \draw[arr] (l1b) -- (l2a);
\draw[arr] (l1c) -- (l2b); \draw[arr] (l1d) -- (l2b);
\node[op] (l3a) at (0.45,-1.95) {$+$};
\draw[arr] (c)   -- (l3a); \draw[arr] (l2a) -- (l3a);
\node[op] (fin) at (1.4625,-2.6) {$+$};
\draw[arr] (l3a) -- (fin); \draw[arr] (l2b) -- (fin);
\end{tikzpicture}
\caption{Group pairwise}
\end{subfigure}
 
\vspace{0.8em}
 
\begin{subfigure}[b]{0.48\linewidth}
\centering
\begin{tikzpicture}
\path (0,-2);
\node[inp] (c)  at (0,0)    {$c$};
\node[inp] (p0) at (0.45,0)  {$p_0$};
\node[inp] (p1) at (0.9,0)   {$p_1$};
\node[inp] (p2) at (1.35,0)  {$p_2$};
\node[inp] (p3) at (1.8,0)   {$p_3$};
\node[inp] (p4) at (2.25,0)  {$p_4$};
\node[inp] (p5) at (2.7,0)   {$p_5$};
\node[inp] (p6) at (3.15,0)  {$p_6$};
\node[op]  (fa) at (1.575,-1.8) {$+$};
\draw[arr] (c)  -- (fa); \draw[arr] (p0) -- (fa);
\draw[arr] (p1) -- (fa); \draw[arr] (p2) -- (fa); \draw[arr] (p3) -- (fa);
\draw[arr] (p4) -- (fa); \draw[arr] (p5) -- (fa);
\draw[arr] (p6) -- (fa);
\end{tikzpicture}
\caption{Fused, e.g.\ a single block FMA}
\end{subfigure}%
\hfill
\begin{subfigure}[b]{0.48\linewidth}
\centering
\begin{tikzpicture}
\node[inp] (c)  at (0,0)    {$c$};
\foreach \i in {0,1,2,3,4,5,6,7} {
  \pgfmathsetmacro{\x}{0.44 + \i*0.44}
  \node[inp] (p\i) at (\x,0) {$p_\i$};
}
\node[op] (fa1) at (0.88,-0.9) {$+$};
\draw[arr] (c)  -- (fa1); \draw[arr] (p0) -- (fa1);
\draw[arr] (p1) -- (fa1); \draw[arr] (p2) -- (fa1); \draw[arr] (p3) -- (fa1);
\node[op] (fa2) at (2.42,-1.8) {$+$};
\draw[arr] (fa1) -- (fa2); \draw[arr] (p4) -- (fa2);
\draw[arr] (p5) -- (fa2); \draw[arr] (p6) -- (fa2); \draw[arr] (p7) -- (fa2);
\end{tikzpicture}
\caption{Chain of fused, e.g.\ MMA or reduction across tensor cores}
\end{subfigure}
\caption{Four summation topologies that compute $c + \sum_i p_i$. All produce the same result in exact arithmetic, but not in general under floating-point rounding. Adapted from~\cite{xie2025mmasim_v1}.}
\label{fig:summation_trees}
\end{figure}
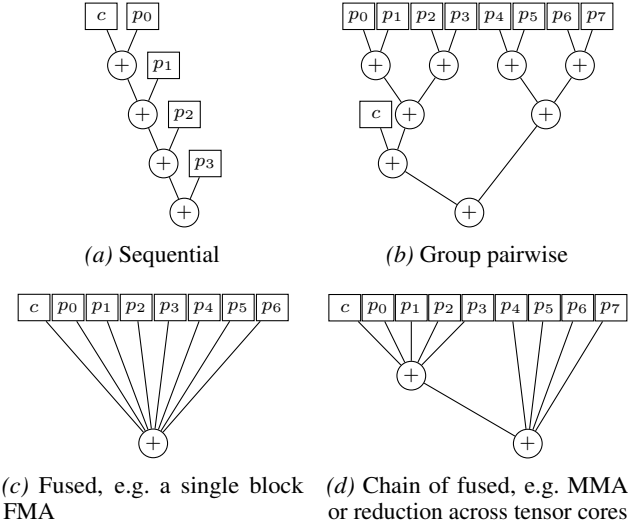

On a GPU computing a forward pass of a FFN or attention block in a transformer, there are two distinct factors deciding reduction orders of element summation (e.g.\ when computing a matrix element of a GEMM\footnote{General Matrix Multiply, a fundamental operation dominating most FLOPs of AI inference. Defined as $C \leftarrow \alpha AB + \beta C$~\cite{nvidia_gemm_perf}.}):

\begin{enumerate}[topsep=0pt, itemsep=0.3em, parsep=0pt, partopsep=0pt]
    \item \textbf{Hardware-level reduction:} An NVIDIA GPU performs 
    GEMMs in subdivided tiles, with the smallest sub-tiles allocated to 
    tensor cores~\cite{nvidia_ptx_isa}. Tensor cores perform matrix 
    multiply-accumulate (MMA) via chains 
    (Figure~\ref{fig:summation_trees}(d)) of fused block multiply-add 
    (block FMA) units. The topologies of both are 
    \textit{fixed in silicon} (i.e.\ the number of p-elements in block 
    FMA as shown in Figure~\ref{fig:summation_trees}(c) and the chaining 
    inside an MMA)~\cite{khattak2025tensor, 
    xie2026mmasim_v2}\footnote{\label{fn:mma}For instance, on the A100 BF16 tensor core 
    (SM80), each block FMA sums 8 products with the accumulator~$c$. 
    One MMA instruction (\texttt{mma.sync.aligned.m16n8k16}) chains 
    two block FMAs to cover its $K=16$ reduction 
    depth~\cite{khattak2025tensor, xie2026mmasim_v2}.}. Software can 
    chain many MMA operations together, but cannot change a single 
    MMA's reduction order\footnote{There are variants of MMA using 
    different chaining of block FMAs, but any individual MMA 
    instruction has a fixed reduction tree.}. CUDA cores, in contrast, are 
    scalar Arithmetic Logic Units (ALUs) without internal reduction. 
    Any multi-operand reduction on CUDA cores is orchestrated by 
    software instructions (warp shuffles, shared-memory trees).
    \item \textbf{Software-level reduction:} Software decides the 
    reduction order at every level above the MMA. Here we distinguish 
    two cases:
    \begin{enumerate}[topsep=0.3em, itemsep=0.3em, parsep=0pt, label=(\alph*)]
        \item \textbf{Static:} Kernel dispatch follows deterministic 
        heuristics that depend on tensor shapes, data types, and library 
        version. The reduction order is thus fully determined by the 
        software stack and input shapes, not by runtime 
        scheduling~\cite{thinkingmachines2024}.
        \item \textbf{Atomic:} Some kernels reduce partial results via 
        \texttt{atomicAdd} on floating-point outputs, where the 
        accumulation order depends on the GPU's warp scheduler and is 
        not reproducible across runs\footnote{The warp scheduler is 
        non-deterministic because it makes dispatch decisions based on 
        runtime state that varies between runs: which warps are 
        stalled on memory~\cite{nvidia_nsight_compute}, cache state 
        and clock frequency (which NVIDIA itself controls for 
        profiler-reproducibility by purging caches and adjusting 
        clocks~\cite{nvidia_nsight_compute}), thermal throttling 
        under sustained load~\cite{jia2019turing}, and 
        memory-controller request reordering across 
        warps~\cite{jog2014dramdivergence}. Two identical kernel 
        launches on the same GPU will have different warps ready at 
        different cycles, so when multiple warps race to \texttt{atomicAdd} to the same output address, the order they arrive is effectively random~\cite{shanmugavelu2024fpna, thinkingmachines2024}.}. This is the sole source of genuine non-determinism we identified in our experiments, and was limited to specific INT de-quantization kernels (see Section~\ref{sec:empirical}). 
    \end{enumerate}
\end{enumerate}

On identical hardware, apparent non-determinism arises when reduction orders do not match. This is the case when using different (software version-dependent) kernel libraries or batch sizes (which change the tensor shapes, in turn triggering different kernel selection and launch-parameter choices such as tile sizes and split-K factor). On different hardware, additional factors include different MMA topologies in different tensor core generations, as well as different approximations used in element-wise calculation on special function units.

\section{Empirical Verification of Determinism}
\label{sec:empirical}

\subsection{Methodology}
\label{sec:empirical_method}

We generally use unmodified vLLM and HuggingFace Transformers on various NVIDIA GPUs hosted by RunPod and Vast.ai. No determinism flags are set in PyTorch or CUDA. We compute L2 distance between extracted signal vectors (log-probabilities, key vectors and occasionally hidden states for HF transformers, only log-probabilities for vLLM\footnote{The public vLLM API only exposes log-probabilities. Intermediate tensors are technically accessible via PyTorch forward hooks on the internal model, but this is fragile across versions, incompatible with CUDA graph capture, and does not straightforwardly reconstruct the full KV cache under tensor parallelism.}) across multiple reference prompts, constructed from natural language prompts (typically at 10,000 token sequence length, but also with occasional stress-tests at longer sequences). When comparing decode inference outputs across conditions (across hardware, batch sizes, etc.), we used teacher-forcing to ensure computation ran on identical tokens.

\subsection{Results}
\label{sec:empirical_results}

\paragraph{Reproducibility check within fixed conditions}
We ran prefill and decode inference with identical models and inputs and found bitwise identical results when fixing the following conditions ($L2 = 0$): hardware SKU; software stack (CUDA toolkit, PyTorch, inference engine and version, attention backend); quantization format and kernel variant; and tensor parallelism rank. When varying batch size, we left the first batch element's inputs unchanged and compared how element 0 is affected by batch neighbours. Different physical cards of the same SKU (e.g.\ A100 80\,GB on RunPod (PCIe) vs. Vast.ai (SXM)) produce identical outputs, confirming that variance originates in software and tensor core layouts, not hardware manufacturing defects\footnote{The full GH100 and GA100 dies contain 144 and 128 SMs respectively, while the H100 SXM5 and A100 SXM4 products ship with 132 and 108 active SMs~\cite{nvidia2022hopper, nvidia2020ampere}. Disabling defective cores to match a lower advertised count is standard yield-management practice across CPU and GPU vendors, including explicitly for the H100~\cite{cerebras2025yield}. The active SM count is fixed per SKU, and tensor-core reduction trees are fixed in silicon~\cite{khattak2025tensor, xie2026mmasim_v2} Bit-exactness across physical cards of the same SKU is therefore unsurprising}.

When testing models of different architectures and quantization formats, (including Mistral, Qwen, Kimi, Deepseek and GLM models at BF16, FP8, INT8 and INT4), we encountered genuine non-determinism for some, but not all INT-quantized models. Experimentally, we isolated the root cause to specific INT-dequantization kernels: The same model, Qwen 3 8B, produced deterministic outputs for AWQ quantized weights, and GPTQ quantization when activating vLLM's \texttt{marlin} kernel. When deactivating said kernel for GPTQ, log-probabilities of repeated prefill and decode runs diverged. Deactivating \texttt{marlin} makes vLLM default to the \texttt{exllama} kernel inside \texttt{$vllm/csrc/quantization/gptq/q\_gemm.cu$}. For example:

\verb|//q_gemm.cu:319-320|
\verb|atomicAdd(out,result01);|
\verb|atomicAdd(out + 1, result23);|

With other occasions of atomics in that same kernel. We conclude that our experimental measurements confirm that inference on NVIDIA GPUs using vLLM and HF Transformers kernels is deterministic unless atomic functions are called in the backend (which has become rare for modern inference engines).

\paragraph{Comparison across conditions}

We found outputs to be invariant for some conditions. Pipeline parallelism rank (tested with Mistral Small 3 spread over 1, 2 and 4 A100 GPUs) is an unsurprising example: GPUs process a transformer forward pass layer by layer, and transmission of activations from one GPU to the next is deterministic data transfer. Minor firmware differences also left outputs invariant, while major kernel library updates created 1-5\% divergence in hidden states (L2 distance/magnitude)~\footnote{The version bump cu128 to cu129 did not introduce A100-targeted kernel updates to cuBLAS, and inference results were unaffected. We did however measure differences across cu118 and cu12.0}. 

We observed sequence-length-dependent occasions of "equivalence classes" in outputs in vLLM and Transformers: While adding batch neighbors generally altered batch element 0's outputs, with increasing sequence length the differences in outputs across batch sizes began to disappear. This effect was most pronounced in prefill inference, where differences disappeared at larger batch sizes first, with complete batch size invariance beginning at sequence lengths above 4000 tokens. For decode inference, we observed a similar effect: occasionally, adjacent batch sizes left batch element 0 unaffected, (e.g. batch size 4 vs. batch size 5), while increasing sequence length made such invariant groups rarer. In no experiments did the \textit{token identities} of batch neighbours affect the first batch element in any way.

\begin{figure}
\centering
\centerline{\includegraphics[width=\columnwidth]{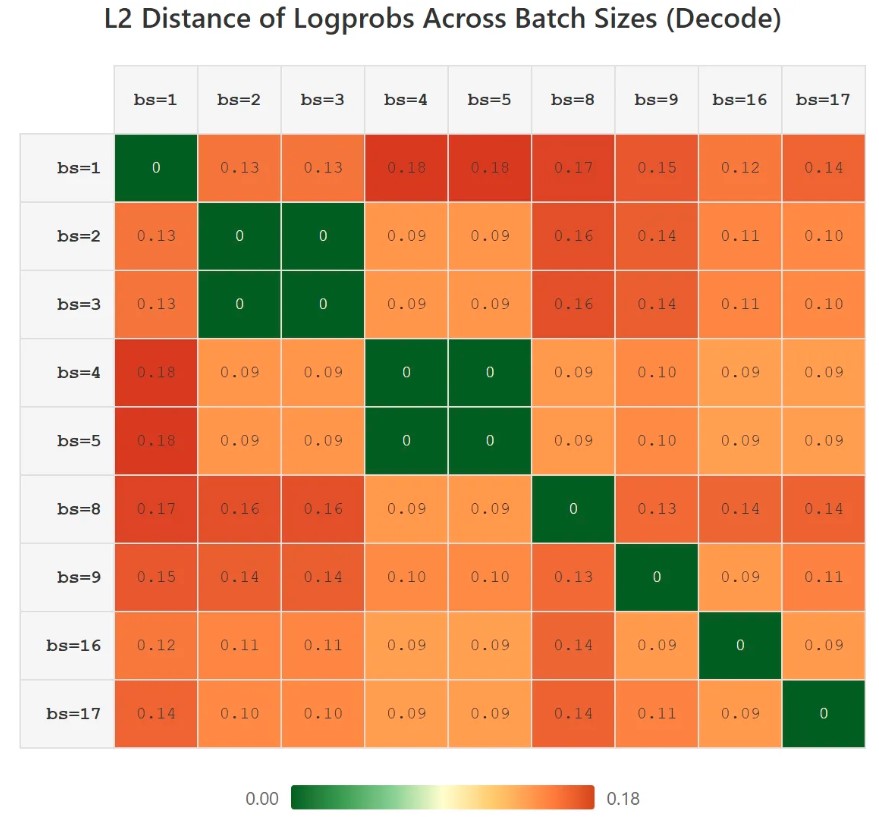}}
\caption{
  Comparison of top-5 logprobs (average across measurements at the last three token positions of the first batch element) across different batch sizes for decode inference in vLLM. We see that some changes in batch size can leave batch element 0's outputs unchanged. Such "equivalence classes" were consistent across prompts for any given fixed sequence length, but they became rarer with increasing sequence length. 
}
\label{fig:batches}
\end{figure}

These results are the product of the interaction of kernels with input tensor shape: In prefill inference, the forward pass behaves as a large matrix multiplication (GEMM) where the row dimension is $\text{M} = \text{Batch} \times \text{Sequence Length}$. For small sequences, the kernel dispatcher employs different tiling strategies and Split-K variants to utilize the GPU, altering the reduction tree as batch size changes. As sequence length increases, we empirically observe that kernel choices stabilize, leading to invariant outputs across batch sizes for sufficiently long sequences (e.g.\ above $\sim$4000 tokens in our experiments). Decode inference behaves as a memory-bound matrix-vector operation (GEMV) where $M = \text{Batch Size}$. In this regime, implementations favor specific dimension alignments for efficient execution (e.g.\ multiples compatible with Tensor Core usage), which can lead to identical reduction trees for groups of batch sizes. However, as the KV-cache (and thus the attention reduction dimension) grows with sequence length, kernel selection and tiling may again change, gradually reducing these equivalence classes. Deepseek-AI independently reports on this issue in the technical report of their V4 model family: \textit{"Traditional cuBLAS library cannot achieve batch invariance"}.~\cite{deepseek2026v4}

Lastly, we examined the relative contribution of different sources of discrepancy and observed that rounding differences from distinct factors (e.g., hardware SKU, kernel implementation, and algorithmic choices) combine in an approximately additive manner in practice. This was particularly apparent during development of the emulator (Section~\ref{sec:emulator}), where incrementally correcting mismatches between the simulation and ground truth consistently reduced the remaining error. Empirically, we observe a rough hierarchy in the magnitude of these effects: quantization format and numerical precision have the largest impact, followed by the attention implementation (e.g.\ SDPA, eager, FlashAttention), hardware SKU, and then remaining factors such as inference mode (prefill vs.\ decode), batch size, and tensor-parallel configuration.

\section{Bit-Exact Emulation of GPU-Accelerated LLM Inference}
\label{sec:emulator}

We built a software-only C and Python emulator capable of predicting every bit of intermediate tensors in a transformer forward pass across multiple NVIDIA GPU architectures. In most tests, we emulate individual FFN and ATTN blocks of Qwen 3 4B in isolation, but we also validated a full forward pass through the model: token embedding, all 36 transformer blocks in series (attention + FFN), final RMSNorm, and the language-model head. Our results are presented in Appendix~\ref{app:emulator-table}.

\subsection{Tensor Core Arithmetic Model}
\label{tc-model}
We emulate the tensor core's block FMA operation and the PTX-instructed MMA (see footnote ~\ref{fn:mma}). Following the hardware characterizations by \citet{khattak2025tensor} and \citet{xie2026mmasim_v2}, block FMA computes raw products without FP32 normalization. Instead, products are aligned to a maximum exponent within a fixed-point window (e.g., 26 bits for the A100: 2 integer, 23 fraction, 1 extra alignment bit), bits shifted out of the window are truncated, the components are summed as integers, and the final result is truncated to FP32. Our emulator replicates this exact arithmetic, with tensor core profiles for each SKU (we created profiles for Lovelace, Ampere, Hopper and Blackwell SKUs based on prior work by \citet{khattak2025tensor} and \citet{xie2026mmasim_v2}).

\subsection{Special Function Units}
\label{SFU}
NVIDIA GPUs accelerate transcendental functions via Multi-Function Units (MUFU). Instructions like \texttt{MUFU.RSQ} (reciprocal square root), \texttt{MUFU.EX2} (exponential with base 2), and \texttt{MUFU.RCP} (reciprocal) rely on architecture-specific silicon lookup tables. We found their outputs are deterministic but deviate from IEEE-correct\cite{ieee754_2019} rounding by up to $\pm 2$ ULP. By exhaustively probing these instructions via inline PTX, we cached their exact hardware outputs (e.g.\ a 4GB table mapping every 32-bit input for \texttt{MUFU.EX2}), ensuring bit-exact SFU emulation without needing GPU access at verification time.

\subsection{Software Reduction, Kernel and RoPE Emulation}
\label{kernels}
Matching GPU-accelerated outputs also requires meticulously modeling kernel-specific reduction trees and compiler optimizations:
\begin{itemize}[topsep=0pt, itemsep=0pt, parsep=0pt, leftmargin=*]
    \item \textbf{GEMM:} Most FLOPs in transformer inference are in matrix multiplication, and above the tile sizes handled fully within tensor cores, software decides tile boundaries and reduction ordering. To achieve bit-exact emulation, our code replicates the exact accumulation path dictated by the chosen kernel configuration:
    \begin{itemize}
        \item \textbf{K-Iteration Order:} The intermediate value of the FP32 accumulator dictates the maximum exponent used for the hardware's fixed-point alignment window. Therefore, the exact sequence in which blocks are added is mathematically load-bearing. Our emulator mirrors the sequential K-walk used by the target kernel's mainloop (CUTLASS's default configuration).

        \item\textbf{BF16 epilogue:} After each projection's FP32 accumulation, the GPU stores the result to global memory as BF16. Our emulator mirrors this by casting the raw FP32 accumulator to BF16 before it enters the next stage. We verify both the raw FP32 accumulator and the BF16-cast output against CUTLASS to confirm that neither the accumulator nor the epilogue hides a discrepancy.
        
        \item \textbf{Live Accumulator State (Fused Kernels):} Standard projections typically initialize the accumulator (depicted as ''c'' in figure \ref{fig:summation_trees}) to zero. However, in fused operations like FlashAttention-2, the PV matrix multiplication accumulates directly into a live, running register ($O_{acc}$) across KV tiles. Our emulator addresses this using a specialized \texttt{block\_fma\_batch} primitive that applies the hardware block FMA to an existing accumulator state,  matching the alignment window shifts that occur on hardware.
        
    \end{itemize}
    \item \textbf{Memory Boundaries:} When the GPU writes BF16 tensors to global memory between pipeline stages (e.g., after RoPE and before FA2), we explicitly enforce a BF16 quantization boundary in the emulator to ensure the tensor core receives the exact same significands as the physical hardware.
    \item \textbf{RMSNorm:} We emulate PyTorch's parallel \texttt{reduce\_kernel} (including its warp-shuffle topology, which changes across PyTorch versions), the \texttt{nvcc} compiler's optimization of division into multiply-by-reciprocal, the \texttt{MUFU.RSQ} rounding, and the specific cast ordering (FP32 normalization $\to$ BF16 cast $\to$ BF16 weight multiply).
    \item \textbf{FlashAttention-2:} FA2 (2.8.3) emulation required resolving several nuanced interactions between the hardware and the compiler's optimizations:
    \begin{itemize}
        \item \textbf{Accumulator Initialization:} As explained above under ''GEMM''.
        \item \textbf{Online Softmax and SFU Usage:} We emulate FA2's block-wise execution pattern: FA2 maintains running statistics for the maximum score $m^{(i)}$ and the sum of exponentials $l^{(i)}$ for each block $i$. When transitioning to a new block, previous running values are scaled by $2^{(m^{(i-1)} - m^{(i)})}$. Our emulator computes these scale factors and the attention weights $P = 2^{S - m^{(i)}}$, and final normalization using our probed MUFU hardware models.
        \item \textbf{Compiler FMA Fusion:} The \texttt{nvcc} compiler fuses operations across inline function boundaries. Specifically, the $l^{(i)}$ rescale multiplication is fused with the addition of the first element of $P$ into a single hardware FMA (FFMA) instruction, resulting in one rounding instead of two. Our emulator mimics this single-rounding step using float64 intermediates.
    \end{itemize}
    \item \textbf{RoPE:} We enforce strictly matching the GPU's CUDA \texttt{libm} \texttt{cosf}/\texttt{sinf} (which differ slightly from CPU \texttt{glibc}) and snap values to BF16 precisely at every stage boundary including QK-norm, RoPE, and FA2.
\end{itemize}

To extend emulation from an initial CUTLASS-target to the proprietary cuBLAS library, we build a one-time per-SKU \textit{dispatch catalog} via cuBLASLt's introspection API, mapping each matmul shape (M,N,K,dtype,layout) to the exact kernel ID cuBLAS dispatches. The search space for any particular model can be substantially narrowed by iterating the weight tensor shapes over the sequence length dimension, which takes minutes and is a one-time setup. For each cataloged kernel, the emulator maps to one of the four reduction topologies that we could narrow down. \footnote{We found no obvious pattern in the way cuBLAS dispatches kernels based on tensor shapes, so we continue to rely on cataloging. For the three tensor shapes of the FFN weight matrices in the Qwen 3 4B model, a catalog of hundreds of thousands of possible tensor shapes is only megabytes of INT indices.}  

\subsection{Diagnostics and Results}
\label{emulatordiag}
We tested our emulator on LLM activations (Qwen3-4B) against GPU ground truth computed using CUTLASS kernel libraries and ---via the dispatch catalog described in Section~\ref{kernels} --- against cuBLAS, which is used by standard Pytorch. Our emulation models individual FFN and ATTN blocks of the transformer stack and can be stacked to simulate a full forward pass.

For the full FFN block (RMSNorm, three matmul projections, SiLU, and residual add), we achieved exactly 0 BF16 diffs (and 0 FP32 accumulator diffs) across all intermediate elements on A100, L40, L40S and H100. For the (significantly more complex) attention block featuring FlashAttention-2, we achieved 0 BF16 diffs (and 0 FP32 accumulator diffs) out of ~71M elements (projections + QK-norm + FA2 scores), at a sequence length of 4,000 tokens. By ''elements'', we mean intermediate tensors captured at multiple in-between states: 

For the FFN block we captured and compared six specific intermediate operations: the raw outputs of the \texttt{gate} and \texttt{up} projections, the element-wise SiLU activation applied to the gate output, the element-wise multiplication of the activated gate and the up projection (\texttt{SiLU(gate) * up}), the output of the \texttt{down} projection, and the final FFN block output following the residual addition. For the ATTN block, we captured and compared the pre-attention RMSNorm output, the individual Query, Key, and Value (\texttt{Q}, \texttt{K}, \texttt{V}) matrix projections, the per-head QK-normalization outputs, the core FlashAttention-2 outputs (which encompasses the Rotary Positional Embeddings and the fused online softmax with QK/PV matmuls), the final Output (\texttt{O}) projection, and the overall attention block output after the residual connection.

Our three-way diagnostic confirmed that the emulator perfectly matches the CUTLASS intermediates with zero differences at all tested sequence lengths (from 64 up to 8000, see Appendix~\ref{app:emulator-table}). We found that CUTLASS-generated outputs diverge from a forward pass using cuBLAS kernels, particularly at short sequence lengths. In the attention block at 4{,}000 tokens, cuBLAS and CUTLASS converge on identical outputs across all pre- and post-attention projections (\texttt{q}, \texttt{k}, \texttt{v}, \texttt{o}), so the CUTLASS-target emulator already matches cuBLAS here without the dispatch catalog being used.

\section{Limitations}
\label{sec:limitations}

\paragraph{MoE inference.}
Our experiments discussed in Section~\ref{sec:empirical_results} include Mixture-of-Experts models such as Qwen3-30B-A3B, GLM-4.6 and Deepseek-v2-coder-lite under tensor parallelism at $\sim$120k-token contexts, all of which produce bit-identical log-probabilities across repeated vLLM runs\footnote{Kimi-K2-Thinking was the sole non-deterministic model in our tests, and we traced this to its INT4 de-quantization, see Section~\ref{sec:empirical_results}}. However, our emulator targets dense FFN and attention blocks, whereas most frontier models are MoE. Beyond the experimental results, we expect emulation to be generalizable to MoE for a more direct reason: Direct inspection of the HuggingFace \texttt{Qwen3MoeSparseMoeBlock} confirms that an MoE extension would require no numerical primitives not already included in our model. For details, see the Appendix ~\ref{app:moe}. We also note that DeepSeek-V4 demonstrates end-to-end batch-invariant and deterministic MoE inference at frontier scale~\cite{deepseek2026v4}.

\paragraph{Only NVIDIA GPUs.} While our experiments were limited to NVIDIA's technology stack, the methodology and general principles transfer directly to other hardware. ~\cite{khattak2025tensor} and ~\cite{xie2026mmasim_v2} have profiled AMD's matrix cores (CDNA2 and CDNA3, i.e., MI200- and MI300-series Instinct GPUs) in a similar manner as they did for NVIDIA's tensor cores, so we expect that further emulation work can demonstrate the same end-to-end inference verification that we applied to NVIDIA's GPUs, CUDA and CUTLASS.

\paragraph{The \texttt{nvjet} kernel family on Hopper.} We found that on Hopper GPUs and sequence lengths below 250 tokens in particular, cuBLAS can occasionally dispatch to deterministic, but proprietary kernels from the \texttt{nvjet} family. This is an edge case that vanishes for longer sequences, and one that could in principle be profiled via black-box methods such as those already used to find MMA reduction order in tensor cores~\cite{khattak2025tensor, xie2026mmasim_v2}.

\paragraph{Backward pass and gradient accumulation in LLM training.} Our experimental and emulation scope was inference only. Training makes use of backward pass kernels and all-reduce, which can introduce atomic functions at multiple levels. FlashAttention-3's Hopper backward accumulates $dQ$, $dK$ and $dV$ via \texttt{atomicAdd} on the non-TMA code paths~\cite{shah2024fa3}\footnote{\texttt{atomicAdd} on $dQ$ at lines 964, 983, 992 of \texttt{hopper/mainloop\_bwd\_sm90\_tma\_gmma\_ws.hpp}; on $dV$ at line 480 and $dK$ at line 511 of \texttt{hopper/epilogue\_bwd.hpp}. A \texttt{Deterministic} template parameter in both files gates per-block semaphore serialization (\texttt{Barrier::arrive\_inc}/\texttt{wait\_eq}) around these accumulations.} and fused normalization backward kernels use atomics for weight gradients. There are deterministic alternatives to atomics, and they are already preferred in frontier LLM training for better reproducibility and stability~\cite{deepseek2026v4}.

\section{Implications for AI Governance}
\label{sec:governance}
We showed that for any (NVIDIA) hardware and software setup used in inference \textbf{that makes no use of floating-point atomic functions}, outputs can be bitwise reproducible without performance-degrading settings if the following information is recorded for later reference\footnote{We only list information related to reduction ordering and hardware-specific element-wise operations, not software-sampling such as temperature}:

\begin{enumerate}[topsep=2pt, itemsep=1pt, parsep=0pt]
    \item \textbf{Hardware SKU} (determines tensor core arithmetic and special function unit behavior)
    \item \textbf{Exact model weights} (in the deployed quantization format)
    \item \textbf{Parallelism topology} (separately for prefill and decode stages)
    \item \textbf{Software versions and any custom kernels} (CUDA toolkit, PyTorch, attention backend, quantization library)
    \item \textbf{Batch size at each forward pass} (one integer per decode step under continuous batching\footnote{If prefill and decode are mixed within a single forward pass (e.g.\ under continuous batching without prefill-decode disaggregation), the per-entry prefill/decode status and sequence lengths must also be recorded, as these determine the kernel-facing tensor shape.}, or record whole batches rather than single sequences)
\end{enumerate}

The verifier can then reconstruct the prover's outputs, or by using identical hardware, by using custom accelerators with identical hardware MMA reduction trees and SFUs, or by emulating all rounding decisions in software.

A caveat applies to the last item. Recording concurrent batch size is only necessary because cuBLAS and standard inference engines are not batch-invariant. If the prover uses batch-invariant kernels such as those deployed in DeepSeek-V4~\cite{deepseek2026v4}, batch size need not be recorded. However, this also makes batch size undetectable \textit{in outputs alone}. More generally, a prover may choose (or be obligated to use) kernels that are not invariant to a governance-relevant property, precisely so that property becomes numerically verifiable.

\paragraph{Cost of re-computation.} Sparse random sampling makes even software-only emulation with no hardware-acceleration viable. The verifier's statistical confidence $P_{detect}(\geq1)$ for an upper-bound percentage $p_{false}$ of misreported records depends only on the total number of samples $k$, not the total number of records to sample from (if this number is $>>k$).
\begin{equation}
\label{eq:detection}
P_{detect}(\geq 1) = 1 - (1 - p_{false})^k
\end{equation}
This works against a covert adversary that attempts to misreport a \textit{percentage} of  computation in large-scale AI inference. Verifying enough samples to provide strong statistical assurance requires only tens to hundreds of CPUs running continuously (see Appendix~\ref{app:flop-calc}), regardless of the prover's total throughput.

For precise verification, the prover need not store all activation tensors, which would be infeasible. A cryptographic hash (or a projection as used by~\cite{karvonen2025difr}) of a single tensor deep in the computational chain --- such as the hidden state at the last layer and last token position of a long sequence --- serves as a highly collision-resistant fingerprint for the entire upstream computation. Bit-exact emulation lets the verifier recompute the expected hash and compare.

Bit-exact reproducibility transforms non-associative rounding from an obstacle to verification into a lever for technical governance capable of identifying the hardware and software stack used for inference. Any \textit{numerically relevant} modifications of kernel libraries, batch composition, and model parallelism can be detected directly in the outputs.

\bibliography{references}
\bibliographystyle{icml2026}

\newpage
\appendix
\onecolumn

\section{CUDA Version Comparison}
\label{app:CUDAversion}
RunPod and Vast.ai A100 instances running CUDA 12.8 and 12.9 (cuBLAS 12.8.3 vs 12.9.0) produced bit-identical hidden states and key vectors across all 29 sampled layers. The cuBLAS changelog for this update lists no additions targeting A100 kernel paths~\cite{nvidia_cuda129_release_notes2}, leaving kernel paths unchanged, and the bit-exact results confirm this empirically. However, when comparing between CUDA 11.8 and 12.1, we found ~1-5\% relative error (L2 distance/magnitude) in hidden states, propagating and accumulating through the model's depth (Qwen 2.5 7B). The cuBLAS 12.0 release notes document kernel-catalog changes (removal of matmul stage constants, Hopper-specific kernel additions that alter dispatch heuristics, and a bias-gradient correctness fix on Ampere), which we identify as the root cause for the change in kernel selection for A100 GEMMs.

\section{MoE Router Analysis}
\label{app:moe}
We inspected the Qwen3 MoE router code to verify that it introduces no new numerical primitives beyond those already modeled.

\begin{itemize}[topsep=2pt, itemsep=0pt, parsep=0pt, leftmargin=*]
    \item The router is \texttt{F.linear} $\to$ FP32 \texttt{softmax}
    $\to$ \texttt{torch.topk} $\to$ $\ell_1$-normalization of the
    top-$k$ weights. Each component is a strict subset of machinery the
    emulator already provides: the linear is a GEMM
    (Section~\ref{kernels}); the softmax uses \texttt{MUFU.EX2} and
    \texttt{MUFU.RCP}, both exhaustively probed
    (Section~\ref{SFU}); \texttt{topk} is an integer sort with no
    floating-point rounding; and the top-$k$ normalization
    $x \cdot \mathrm{reciprocal}(\sum x)$ is structurally identical to
    the final divide in RMSNorm.
    \item Each expert is a gated FFN
    (\texttt{gate\_up\_proj} $\to$ SiLU-gate $\to$ \texttt{down\_proj}),
    i.e.\ the exact computation the emulator already reproduces
    bit-exactly (Section~\ref{emulatordiag}).
    \item Per-expert outputs are accumulated into
    \texttt{final\_hidden\_states} via a Python \texttt{for} loop over
    \texttt{expert\_hit}, returned in sorted order by
    \texttt{.nonzero()}. Within each \texttt{index\_add\_} call,
    \texttt{token\_idx} is duplicate-free (\texttt{torch.topk} returns
    distinct experts per token), so no atomic contention occurs; across
    iterations, the reduction order is fixed by the deterministic
    \texttt{nonzero} ordering. vLLM replaces the loop with a fused
    grouped-GEMM kernel, but the per-expert GEMM and weighted sum are
    again strict subsets of existing emulator primitives.
\end{itemize}

\section{Inference FLOP Calculation}
\label{app:flop-calc}
 
\paragraph{Forward-pass FLOP estimate.}
We use the standard approximation $\text{FLOP}_{\text{fwd}} \approx 2 \cdot N_{\text{active}} \cdot T$ for a transformer forward pass~\cite{kaplan2020scaling}, where $N_{\text{active}}$ is the number of active parameters per token and $T$ is the total number of tokens processed. The factor of two comes from the multiply-accumulate operation dominating matrix multiplication. This formula counts activation-weight matmuls in QKV/O projections and FFN blocks; it does not include the activation-activation matmuls inside attention ($QK^\top$ and $PV$), which scale as $\mathcal{O}(L \cdot d_{\text{model}} \cdot N_{\text{context}})$ per token. For short-context workloads the under-count is negligible; the crossover at which attention FLOPs equal FFN FLOPs occurs near $N_{\text{context}} \approx 3\,d_{\text{model}}$, and this is without factoring current trends towards sparse or linear attention.

\paragraph{Verification workload.}
Random-sample verification of $B$ batches, each containing $S$ sequences of average length $T_{\text{seq}}$, requires
\begin{equation*}
\text{FLOP}_{\text{verify}} = 2 \cdot N_{\text{active}} \cdot B \cdot S \cdot T_{\text{seq}}.
\end{equation*}
Published metadata from OpenRouter's 100T-token usage study~\cite{openrouter_stateofai} reports an average prompt length of approximately 5,400 tokens and an average generated response of 600 tokens per request (late 2025). With $B = 50$, $S = 64$, $T_{\text{seq}} = 6{,}000$, and a representative $N_{\text{active}} = 100\text{B}$ mixture-of-experts model, this yields $k = B \cdot S = 3{,}200$ independently verifiable sequences and approximately $3.8$~EFLOP of recomputation. Scaling to $B = 500$ gives $k = 32{,}000$ and $38$~EFLOP. For workloads with substantially longer average sequences, the budget scales linearly in $T_{\text{seq}}$: at $T_{\text{seq}} = 30{,}000$ it becomes $19$~EFLOP ($B=50$) or $192$~EFLOP ($B=500$); at $T_{\text{seq}} = 100{,}000$, $64$~EFLOP or $640$~EFLOP respectively.

\paragraph{Scale of audited workload.}
For context, OpenRouter reports aggregate daily token traffic of approximately $2 \times 10^{12}$ tokens~\cite{openrouter_stateofai}. At $N_{\text{active}} = 100\text{B}$ (overestimate for open-weights models in early 2026) this corresponds to roughly $400$~ZFLOP per day. The $B=50$ verification workload is therefore approximately $10^{-5}$ of daily OpenRouter-scale compute at the measured average sequence length, rising to approximately $1.6 \times 10^{-3}$ at the $B=500$, $T_{\text{seq}} = 100{,}000$ upper estimate.

\paragraph{Detection-probability scaling.}
Applying \cref{eq:detection} to a covert adversary misreporting 0.1\% of records ($p_{\text{false}} = 10^{-3}$):
\begin{equation*}
P_{\text{detect}}(k=3{,}200) \approx 0.96, \qquad P_{\text{detect}}(k=32{,}000) \approx 1 - 10^{-14}.
\end{equation*}

The $10\times$ scale-up from $B=50$ ($k=3{,}200$) to $B=500$ ($k=32{,}000$) does not materially improve detection at 0.1\% misreporting (already saturated), but extends reliable detection into the 0.01\% range (at 96\% confidence, or 0.005\% at 80\% confidence).

\paragraph{CPU hardware requirement.}
Converting the verification-FLOP figures to sustained throughput over a $86{,}400$-second day yields requirements from approximately $44$~TFLOP/s (at $B = 50$, measured average $T_{\text{seq}} = 6{,}000$) to $7.4$~PFLOP/s (at $B = 500$, $T_{\text{seq}} = 100{,}000$). A current-generation AMD EPYC 9965 CPU provides approximately $27$~TFLOP/s FP32 peak~\cite{amd_epyc9965_spec}; at 60\% sustained utilization, the raw-FLOP requirement maps to roughly $3$ CPUs for the grounded estimate and $\approx 275$ CPUs for the pessimistic estimate.

Both numbers likely need to be multiplied by at least $10\times$, factoring in the overhead of emulating reduction trees. Still, even with a conservative estimate of 100B active parameters, we conclude that the resource requirements of even software-only  bitwise inference verification are manageable, thanks to the favourable scaling of random sampling.

\section{Extended Reproducibility Results}
\label{app:fingerprints}

\begin{table}[h]
\centering
\small
\caption{Cross-hardware $L_2$ distance between Qwen2.5-7B-Instruct prefill
  hidden states (FP16, 1134-token prompt, batch size 1, 10 runs per GPU).
  Diagonal entries are within-hardware across 10 repeats and are all
  exactly $0.000000 \pm 0.000000$. Off-diagonal zeros identify SKUs
  sharing a tensor-core generation; non-zeros identify architecture
  boundaries. Source: \texttt{different\_hardware.py}.}
\label{tab:cross-hw}
\begin{tabular}{lcccccccc}
\toprule
          & A100-SXM4 & A100-PCIe & A40 & H100-NVL & H100-PCIe & H200 & L40S \\
\midrule
A100-SXM4 & --    &       &        &        &        &        &        \\
A100-PCIe & \textbf{0.000} & --    &        &        &        &        &        \\
A40       & 0.496 & 0.496 & --     &        &        &        &        \\
H100-NVL  & 0.526 & 0.526 & 0.549  & --     &        &        &        \\
H100-PCIe & 0.526 & 0.526 & 0.549  & \textbf{0.000} & --     &        &        \\
H200      & 0.526 & 0.526 & 0.549  & \textbf{0.000} & \textbf{0.000} & --     &        \\
L40S      & 0.450 & 0.450 & 0.576  & 0.456 & 0.456 & 0.456 & --     \\
\bottomrule
\end{tabular}

\end{table}

The within-SKU zeros (A100-SXM4 $\leftrightarrow$ A100-PCIe) rule out
physical-card variance: yield-binned chips of the same SKU produce
bit-identical outputs. The within-architecture zeros (H100-NVL,
H100-PCIe, H200) show that tensor-core arithmetic is shared across
products built on the same compute capability, regardless of memory
subsystem or product tier. All non-zero off-diagonal distances
correspond to crossing an architecture boundary (Ampere~$\to$~Hopper,
Ampere~$\to$~Ada, etc.). Batch-size-2 and batch-size-4 panels
(omitted for space) show the same pattern with slightly different
magnitudes. Within-hardware statistical noise was $0.0$ for all seven
SKUs at all three batch sizes.

We also investigated the relative magnitude of numerical deviation in a suite of ablations. A particular focus was on the detectability of software-settings when replaying on different hardware, both in prefill and decode.
We measure this as a signal-to-noise ratio:

\begin{equation*}
\text{SNR} = \frac{\|y_{\text{claimed}\,A} - y_{\text{replayed}\,B}\|_2}
                  {\Delta_{hardware}},
\end{equation*}

The denominator is the same-configuration cross-hardware floor averaged over multiple samples, the numerator averages over configuration mismatches.
SNR $\gg 1$ means the mismatch is detectable above hardware noise,SNR $\approx 1$ means it is not.

\begin{table}[h]
\centering
\small
\caption{Cross-hardware detectability of configuration differences.
  All SNR values are aggregated across reference prompts; logprob SNR
  is reported for comparability, since vLLM only exposes logprobs.
  ``A100$\leftrightarrow$H100'' is the mean over both generation
  directions.}
\label{tab:snr-summary}
\begin{tabular}{llccc}
\toprule
Property varied & Test setup & Prefill SNR & Decode SNR & Detectable \\
\midrule
Attention impl.\ (eager/SDPA/FA2) & HF, Qwen2.5-7B, 10k tok, A100$\leftrightarrow$H100 & 10.5$\times$ & 5.0$\times$ & \checkmark \\
Quant.\ format (AWQ vs GPTQ, INT4) & vLLM, Qwen3-8B, A100$\to$H100 & 26.7$\times$ & 21.0$\times$ & \checkmark \\
Marlin kernel toggle (same format) & vLLM, Qwen3-8B, A100$\to$H100 & 0.99$\times$ & 0.95$\times$ & $\times$ \\
Batch size (9 sizes, 1--17) & HF, Qwen2.5-7B, 10k tok, A100$\to$H100 & 1.08$\times$ & 1.03$\times$ & $\times$ \\
Tensor parallelism (1/2/4) & vLLM, Qwen2.5-7B, 10k tok, A100$\to$H100 & 0.90$\times$ & --- & $\times$ \\
Decode batch size via prefill re-exec & HF, Qwen2.5-7B, A100 only & --- & 0.98$\times$ & $\times$ \\
\bottomrule
\end{tabular}
\end{table}

\newpage

\section{Emulator Three-Way Diagnostic}
\label{app:emulator-table}

\begin{table*}[h]
\centering
\small
\caption{BF16 diff counts for the emulator's three-way diagnostic on
  Qwen3-4B, layer 20. Each cell is (Emu~vs~CUTLASS) / (Emu~vs~Model) /
  (CUT~vs~Model). ``Model'' is HuggingFace Transformers forward with
  \texttt{torch.matmul}, which dispatches to cuBLAS. Denominators are
  element counts per stage at the given sequence length. \textbf{0} =
  bit-exact. Percentage figures refer to the fraction of non-identical elements, regardless of the magnitude of deviation (typically by 1-2 ULP). The emulator matches CUTLASS bit-exactly at every tested
  hardware/sequence-length combination. Divergence from cuBLAS depends
  on kernel dispatch: on L40 the \texttt{down\_proj} kernel choice
  differs at both tested sequence lengths, while on A100 and H100 at
  8k tokens cuBLAS and CUTLASS converge.}
\label{tab:emu-diag}
\begin{tabular}{l@{\hspace{4pt}}cccccc}
\toprule
           & A100, 8k             & H100, 256              & H100, 8k             & L40, 256                  & L40, 8k                   & A100 attn, 4k \\
\midrule
RMSNorm out        & 0/20.48M                   & 0/0.66M                    & 0/20.48M                   & 0/0.66M                        & 0/20.48M                        & 0/10.24M \\
gate\_proj         & 0/77.82M / 0 / 0           & 0/2.49M / 0 / 0            & 0/77.82M / 0 / 0           & 0/2.49M / \textbf{37.3\%}      & 0/77.82M / 0 / 0                & -- \\
up\_proj           & 0/77.82M / 0 / 0           & 0/2.49M / 0 / 0            & 0/77.82M / 0 / 0           & 0/2.49M / \textbf{37.0\%}      & 0/77.82M / 0 / 0                & -- \\
SiLU(gate)         & 0/77.82M / 0 / 0           & 0/2.49M / 0 / 0            & 0/77.82M / 0 / 0           & 0/2.49M / \textbf{26.6\%}      & 0/77.82M / 0 / 0                & -- \\
SiLU $\cdot$ up    & 0/77.82M / 0 / 0           & 0/2.49M / 0 / 0            & 0/77.82M / 0 / 0           & 0/2.49M / \textbf{48.2\%}      & 0/77.82M / 0 / 0                & -- \\
down\_proj         & 0/20.48M / 0 / 0           & 0/0.66M / \textbf{0.47\%}  & 0/20.48M / 0 / 0           & 0/0.66M / \textbf{71.0\%}      & 0/20.48M / \textbf{37.6\%}      & -- \\
FFN block out      & 0/20.48M / 0 / 0           & 0/0.66M / \textbf{0.09\%}  & 0/20.48M / 0 / 0           & 0/0.66M / \textbf{42.1\%}      & 0/20.48M / \textbf{16.7\%}      & -- \\
\midrule
Q projection       & --                         & --                         & --                         & --                             & --                              & 0/16.38M / 0 / 0 \\
K projection       & --                         & --                         & --                         & --                             & --                              & 0/4.10M / 0 / 0 \\
V projection       & --                         & --                         & --                         & --                             & --                              & 0/4.10M / 0 / 0 \\
O projection       & --                         & --                         & --                         & --                             & --                              & 0/10.24M / 0 / 0 \\
Q-norm             & --                         & --                         & --                         & --                             & --                              & 0/16.38M / 0 / 0 \\
K-norm             & --                         & --                         & --                         & --                             & --                              & 0/4.10M / 0 / 0 \\
FA2 core           & --                         & --                         & --                         & --                             & --                              & 0/16.38M / 0 / 0 \\
\bottomrule
\end{tabular}
\end{table*}

\begin{table*}[h]
\centering
\caption{Bit‑exactness validation across GPUs and sequence lengths.  
All entries are BF16 diff counts (0 = bit‑exact). Percentage figures refer to the fraction of non-identical elements, regardless of the magnitude of deviation (typically by 1-2 ULP). The only remaining gap is the \texttt{nvjet} kernel family on H100 at seq=100 down\_proj. CC = Confidential Computing mode.}
\label{tab:cross_sku_validation}
\begin{tabular}{lcccccc}
\toprule
GPU & Mode & seq=100 & 250 & 1000 & 4000 & Full model \\
\midrule
A100 (sm\_80)      & CUTLASS & 0 & 0 & 0 & 0 & -- \\
A100               & cuBLAS  & 0 & 0 & 0 & 0 & 0 diffs (seq=32,250) \\
\midrule
L40S (Ada sm\_89)  & CUTLASS & 0 & 0 & 0 & 0 & -- \\
L40S               & cuBLAS  & 0 & 0 & 0 & 0 & -- \\
\midrule
H100 (sm\_90)      & CUTLASS & 0 & 0 & 0 & 0 & -- \\
H100               & cuBLAS  & 37.9\%$^\ast$ & 0 & 0 & 0 & -- \\
H100 (CC)          & cuBLAS  & same$^\ast$   & 0 & 0 & 0 & -- \\
\bottomrule
\addlinespace
\multicolumn{7}{l}{\footnotesize $^\ast$ The 37.9\% diffs occur only in \texttt{down\_proj} at seq=100; the cuBLAS-dispatched kernel is from the proprietary \texttt{nvjet} family.} \\
\end{tabular}
\end{table*}

\end{document}